\PassOptionsToPackage{dvipsnames}{xcolor}
\documentclass[conference]{IEEEtran9pt}
\IEEEoverridecommandlockouts
\usepackage{amsmath,graphicx}
\usepackage{subfig}
\usepackage{booktabs}
\usepackage{arydshln}
\usepackage{comment}
\usepackage{subfig}
\usepackage{tcolorbox}
\usepackage[T1]{fontenc}
\usepackage{inconsolata}
\usepackage[hidelinks]{hyperref}
\usepackage{multirow}

\title{
Speech Retrieval-Augmented Generation without\\ Automatic Speech Recognition
\thanks{Accepted at ICASSP 2025.}
}

\author{
\IEEEauthorblockN{
  Do June Min\IEEEauthorrefmark{1}\IEEEauthorrefmark{2}\thanks{$\dagger$Work done while the author was interning at AWS AI Labs.},
  Karel Mundnich\IEEEauthorrefmark{3},
  Andy Lapastora\IEEEauthorrefmark{3},
  Erfan Soltanmohammadi\IEEEauthorrefmark{3},
  Srikanth Ronanki\IEEEauthorrefmark{3},
  Kyu Han\IEEEauthorrefmark{3}
}
\IEEEauthorblockA{\IEEEauthorrefmark{1}University of Michigan, dojmin@umich.edu}
\IEEEauthorblockA{\IEEEauthorrefmark{3}AWS AI Labs, \{kmundnic, allapast, solterfa, ronanks, kyujhan\}@amazon.com}
}

\begin{document}
%
\maketitle
\begin{abstract}
One common approach for question answering over speech data  
is to first transcribe speech using automatic speech recognition (ASR) and then employ text-based retrieval-augmented generation (RAG) on the transcriptions. While this cascaded pipeline has proven effective in many practical settings, ASR errors can propagate to the retrieval and generation steps. To overcome this limitation, we introduce SpeechRAG, a novel framework designed for open-question answering over spoken data.
Our proposed approach fine-tunes a pre-trained speech encoder into a speech adapter fed into a frozen large language model (LLM)--based retrieval model. By aligning the embedding spaces of text and speech, our speech retriever directly retrieves audio passages from text-based queries, leveraging the retrieval capacity of the frozen text retriever.
Our retrieval experiments on spoken question answering datasets show that direct speech retrieval does not degrade over the text-based baseline, and outperforms the cascaded systems using ASR. For generation, we use a speech language model (SLM) as a generator, conditioned on audio passages rather than transcripts. Without fine-tuning of the SLM, this approach outperforms cascaded text-based models when there is high WER in the transcripts.
\end{abstract}
%

\section{Introduction}
\label{sec:intro}

Retrieval-Augmented Generation (RAG) \cite{lewis2020retrieval} has enabled Large Language Models (LLMs) to generate responses using data not available during any of their training stages. This increases the helpfulness of these models since they can be used as an interface to explore topics released after the training, or to cite sources of information to improve factuality. However, RAG remains mainly used for text-based sources, which may contain images and tables \cite{chen2022murag, joshi2024robust}.

Over recent years, however, we have observed a surge in the creation of unstructured content and information such as audio recordings or videos containing spoken information.
One tool that has the potential to enable efficient search through these expanding audio archives to structure the information is spoken content retrieval \cite{lee2015spokencontentretrieval}.
This method indexes and retrieves passages in audio format and offers a solution for searching large collections of speech data such as meeting recordings \cite{chelba4490200}. 
However, available spoken content retrieval systems treat the problem as a variant of text retrieval, where the source text is automatic speech recognition (ASR) transcriptions of audio produced by an ASR system \cite{lee2015spokencontentretrieval}, which usually contain errors.
This system is often referred to as a \textit{cascaded} model, where the output of the ASR step is fed to the text-based retriever as input.
With the recent advancements in speech processing \cite{radford2022robustspeechrecognitionlargescale} and text retrieval \cite{muennighoff-etal-2023-mteb},
the cascaded model presents a robust baseline across many speech tasks, including spoken dialogue state tracking \cite{jacqmin-etal-2023-olisia} and spoken language understanding (SLU) \cite{serdyuk2018}.

Despite the robustness of the cascaded approach, there are downsides in its use for retrieval from data in spoken form. 
For example, the errors from ASR can propagate downstream, and negatively impact retrieval and generation performance.
This problem can be exacerbated in challenging topics for ASR systems such as named entity recognition \cite{szymanski-etal-2023-arent}, 
even more so considering that named entities are often central to accurate retrieval since often they are used to match queries to passages.
In addition, applying ASR to speech results in the loss of paralinguistic information. 
To best represent the information contained in speech without loss, speech should be indexed and retrieved in its original form.

In this work, we propose a solution to overcome the limitations of the ASR-based cascaded retrieval systems and instead directly index and search audio passages in their original speech format.
Specifically, we tackle the problem of text query to audio passage retrieval by using a text embedding model to
encode audio passages in the same text embedding space, effectively allowing multimodal retrieval using a single embedding model.


Our main contributions are:
\begin{itemize}
\item We propose and implement an end-to-end speech retrieval system that embeds both text and audio in the same space, allowing retrieval of text and speech interchangeably,
\item Our method used a lightweight adapter between an LLM-based text embedding model and a speech encoder, making it data efficient during training and avoiding cross-modal contrastive learning,
\item We implement an end-to-end speech RAG framework that requires no ASR for open question answering from spoken passages.
\end{itemize}

\begin{figure*}%
    \centering
    \subfloat[\centering Speech retriever.]{{\includegraphics[width=0.25\textwidth,height=\textheight,keepaspectratio]{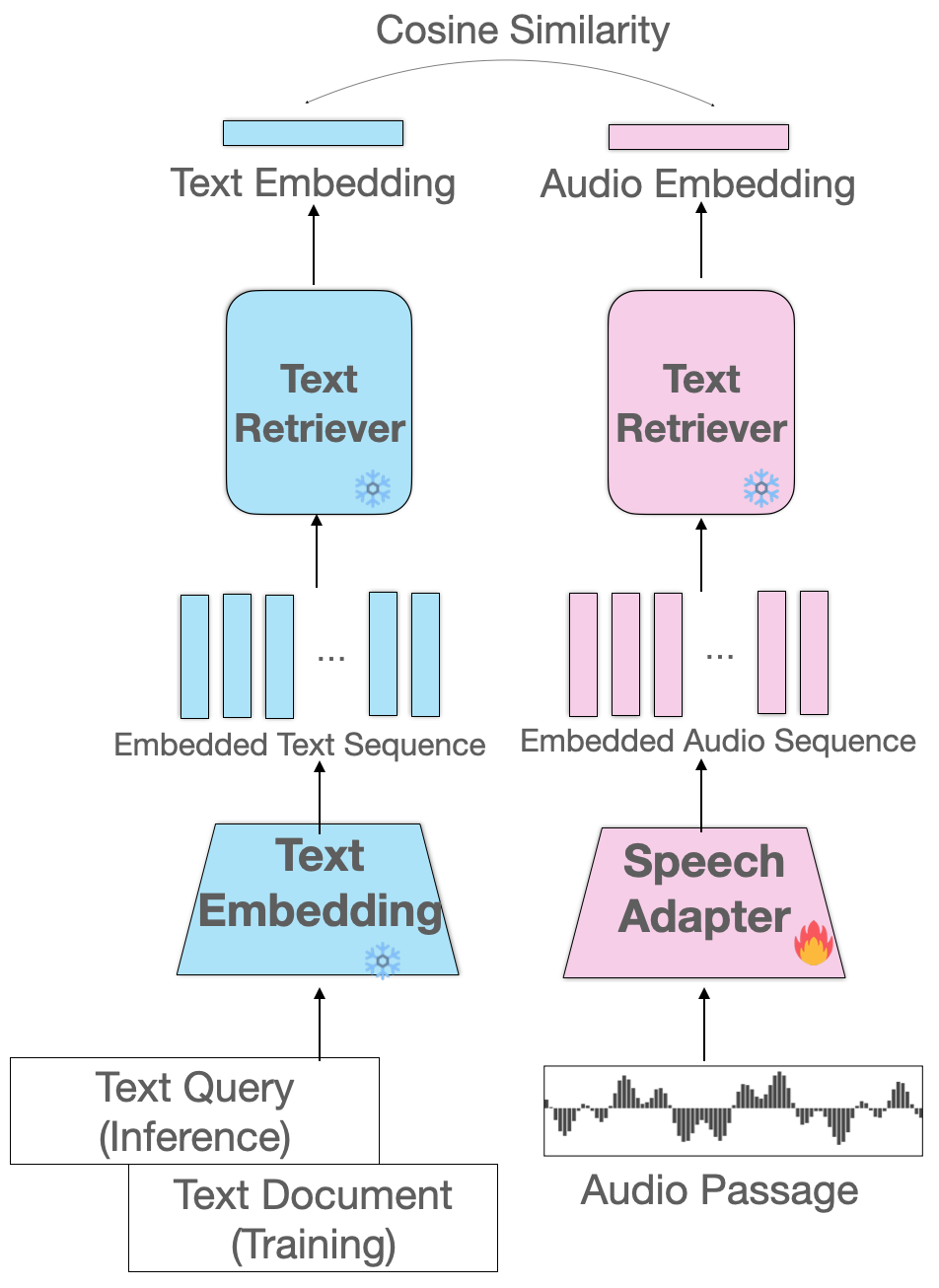} }}%
    \subfloat[\centering Our SpeechRAG framework (top) and cascaded system (bottom).]{{\includegraphics[width=0.68\textwidth,height=\textheight,keepaspectratio]{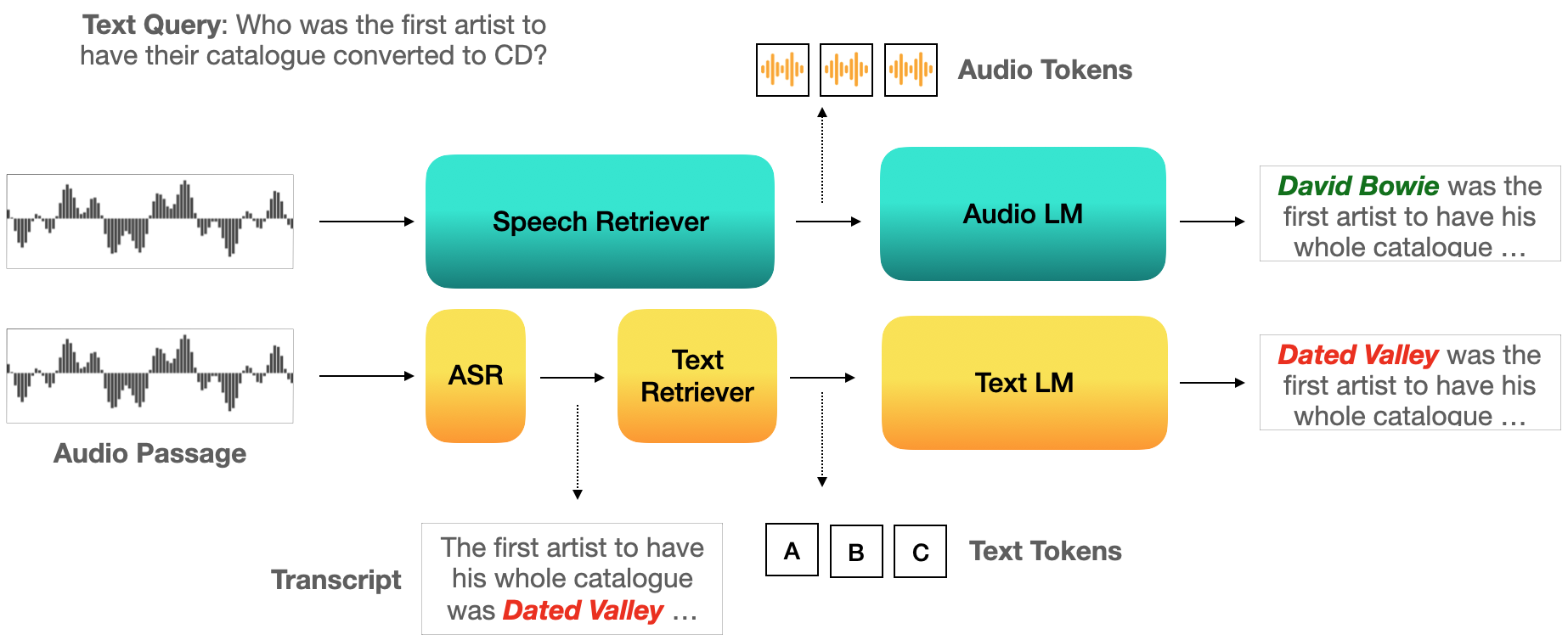} }}%
    \qquad
    \caption{
    \textbf{(a) Our speech retriever} consists of an adapter that projects speech into the embedding space of the text retrieval model. 
    During training, we use distillation from the text embedding of the transcript of the speech to refine our speech embedding.
    This allows us to leverage the frozen text retriever's capacity during a similarity search with the query.
    During testing, we use the text branch (Fig 1a, left) to embed text queries and the speech branch (Fig 1a, right) to embed speech passages.
    \textbf{(b) Our SpeechRAG framework} consists of a speech retriever and an SLM (Figure 1b, top) and operates directly on speech.
    On the other hand, ASR-based cascaded baselines (Fig 1b, bottom) first transcribe the audio and use text-based RAG, leading to the propagation of ASR errors in retrieval and generation.
    }%
    \label{fig:model}%
\end{figure*}

\section{Background}
\label{sec:background}

Inspired by the success of CLIP \cite{Radford2021LearningTV} in the task of cross-modal retrieval, several audio embedding models 
have been proposed, leveraging large audio-caption pair datasets with contrastive learning techniques \cite{
yusong_clap2023, 
CLAP2022, robinson2021contrastivelearninghardnegative, chen2022why}.
Moreover, powerful audio and speech encoders have been made widely available, leading to better audio representations for a wide array of tasks
 \cite{radford2022robustspeechrecognitionlargescale, Duquenne:2023:sonar_arxiv}.
However, models like CLAP-LAION, CLAP-MS, and others are limited to matching audio to natural language descriptions of an audio event,
instead of matching speech to its corresponding language content \cite{oncescu2021audioretrievalnaturallanguage}. 

On the other hand, spoken content retrieval involves matching the linguistic and semantic content of speech contained in audio to queries \cite{lee2015spokencontentretrieval, Chuang2019SpeechBERTCP}.
Recent models typically use contrastive learning \cite{recap2024, Wang2024RetrievalAE} to train speech embedding models that can either match spoken or text queries to audio passages.
SpeechDPR tackles spoken query to spoken passage retrieval by training an end-to-end model with teacher distillation from dense text retrievers \cite{lin2024speechdprendtoendspokenpassage}.
Our work tackles text query to audio passage retrieval and avoids expensive contrastive learning while achieving competitive performance with ground truth (GT) text retrieval.

Finally, another related work is ReSLM \cite{wang2024retrieval}, where the authors propose to use a retrieval approach to boost the ASR performance of an SLM in named entity recognition. However, this approach differs from ours in that the retrieval is performed to obtain different text representations for specific spoken words to augment the prompt used for ASR.

\section{Retrieval from Spoken Passages}
\label{sec:speechrag}

Our framework for text-to-speech retrieval consists of a fine-tuned speech adapter and a frozen 
unimodal text retriever (shown in \autoref{fig:model}a).
The speech adapter projects downsampled speech representations into the text embedding space of the text retriever model.
This architecture allows us to leverage the retrieval capacity of state-of-the-art text retrievers without optimizing a cross-modal
embedding model from scratch, which would require much larger datasets.

\subsection{Speech adapter for text-based retriever}
\label{subsec:speech_adapter}
The function of the speech adapter is to adapt the speech representations of 
the input audio into the text embedding space of the downstream text model.
In this work, we follow a similar approach as \cite{das2024speechverselargescalegeneralizableaudio}.
Specifically, the adapter, which consists of an encoder and a downsampler, is trained 
using 
a cosine embedding similarity loss propagated in an end-to-end manner from the text model.

\noindent\textbf{Speech encoder:}
We use HuBERT \cite{hsu2021hubertselfsupervisedspeechrepresentation}, a pre-trained speech encoder, and feed the last hidden layer representations to the speech adapter.

\noindent\textbf{Speech adapter:}
Typically, the text token sequence length of a speech transcript is much shorter than the sequence length of the corresponding discretized speech. Therefore, we use an average pooling layer of the time dimension to downsample the speech representations, and a projection layer to upsample the speech representation to the LLM embedding dimension \cite{kang2024prompting}.

\subsection{Cross-modal retriever}
\label{subsec:retriever}
To encode both modalities (text for queries and speech for passages), we use the same frozen text-based retriever as our backbone \cite{reimers-2019-sentence-bert, wang2024improvingtextembeddingslarge}.
The main difference across modalities is the embedding module, which maps the raw data (text tokens, audio features)
into the embedding space of the text retriever.

\noindent\textbf{Embedding text:}
We use the original token embedding module of the text retriever model. Then, the input embedding sequences are 
processed by the retriever, and the final layer's representations are pooled to create $e_t$, the fixed-sized representation of the text.

\noindent\textbf{Embedding speech:}
Our speech adapter embeds speech into a sequence of embeddings by first deriving frame-level speech features with the pre-trained speech encoder, and downsampling the feature sequences over the temporal dimension.
The embeddings are then consumed by the retriever, with
$e_s$ as the final embedding output.

\noindent\textbf{Training loss:}
Given $e_t$, the text embedding of a ground truth transcript of an audio passage and
$e_s$, the audio embedding of the audio passage, we compute the distillation loss using the cosine embedding loss:
\begin{equation}
\mathcal{L}(e_s,e_t) =  1 - \cos(e_s, e_t) =  1 - \frac{e_s \cdot e_t}{||e_s||||e_t||}.
\end{equation}

\subsection{Audio-conditioned generator}
\label{subsec:generator}
For an end-to-end speech RAG model that does not require ASR, we use a pre-trained multi-task SLM trained on various speech tasks such as speech recognition or question answering \cite{Qwen-Audio}.
Similar to our cross-modal retriever, the SLM consists of a speech adapter and a frozen text model.


\section{Experiments}
\label{sec:experiments}

\subsection{Data}
\label{subsec:data}
We use two speech datasets,
SpokenSQuAD
\cite{lee2018spoken},
and 
VoxPopuli \cite{wang-etal-2021-voxpopuli} for our RAG experiments.
SpokenSQuAD is a spoken version of the SQuAD dataset, where Wikipedia text are converted into speech 
using text-to-speech systems \cite{rajpurkar-etal-2016-squad}.
It is annotated with text queries about the spoken passages and ground truth answers.
The audio passages are pre-chunked in passage-level, with the average spoken passage duration of $\sim$60s.

VoxPopuli is a large collection of speech from the European Parliament events, with utterances averaging $\sim$10s.
We use the English subset of VoxPopuli and create query and answer pairs from the speech using LLM-prompting.
We first identify potential answer candidates by extracting named entities from each utterance using a fine-tuned BERT model \cite{tjong-kim-sang-de-meulder-2003-introduction}, and use answer-aware generation to mine text queries.
Specifically, we prompt Claude 3.5 Sonnet to generate a question that is answered by the extracted named entity, given the utterance as context (\autoref{tab:data_examples}).
For both datasets, each query has only one relevant passage.
\begin{table}[t]
\caption{Example of a generated query and answer in VoxPopuli.}
\centering
\footnotesize
\resizebox{1.\linewidth}{!}{
\begin{tabular}{lp{0.8\columnwidth}}
\toprule
Passage   & The conclusion of the Framework Agreement provides a legally...\\\midrule
Query     & What legally binding instrument provides for upgrading and strengthening EU-Australia bilateral relations as well 
as increasing cooperation between them? \\\midrule
GT Answer & Framework Agreement                                        \\ \bottomrule                   
\end{tabular}
}
\label{tab:data_examples}
\end{table}

\subsection{Implementation details}
\label{sec:implementation}
\noindent\textbf{Retriever:}
For our speech encoder, we use HuBERT-large \cite{hsu2021hubertselfsupervisedspeechrepresentation}, which uses self-supervised learning to generate deep representations of speech sampled at 16kHz.
Our speech adapter downsamples the output of HuBERT 4 times (for a final frame length of 80ms). Both the speech encoder and adapter are unfrozen.
For our frozen retriever backbone, we use E5-Mistral-7B-Instruct LLM-based retriever 
\cite{wang2024improvingtextembeddingslarge}.
We train the model for 20 epochs, with a stopping criteria of validation loss, with a patience of 3.
We use the Adam optimizer with a learning rate of 5e-5, and $\beta_1=0.9, \beta_2=0.999$.
We train with a batch size of 4 and a gradient accumulation step of 16.

\noindent\textbf{Generator:}
After audios are retrieved from the vector database, each query is combined with the top-$k$ retrieved audios and a task instruction prompt as a prompt
an instruction-tuned LLM \footnote{\url{https://hf.co/mistralai/Mistral-7B-Instruct-v0.2}} which generate answers.
to an 7B-parameter SLM \footnote{\url{https://hf.co/Qwen/Qwen-Audio-Chat}} to generate an answer
 \cite{Qwen-Audio}.
 We use the SLM as is, without fine-tuning its parameters to this specific task.
 
\subsection{Evaluation}
\label{subsec:evaluation}
\noindent\textbf{Retrieval:} We use Recall$@k$ as our evaluation metrics for the retrieval experiment:
\begin{equation}
\text{Recall@}k = \frac{\text{\# of relevant passages in top-$k$}}{\text{Total \# of relevant items}},
\end{equation}
where $k=5,10,100$.
Each query has exactly one relevant passage. For the relevance score, we use the cosine similarity between the text query embedding
and the audio passage embedding.

\noindent\textbf{Generation:}
For the generation experiment, we use top-5 retrieved passages as context provided along 
with the text query and the LLM instruction.
To evaluate the correctness of the generated answer, we use 
\begin{itemize}
\item \textit{\textbf{Exact Match (EM)}}, which assigns $1$ if the ground-truth is in the answer, otherwise $0$, and
\item \textit{\textbf{LLM Correctness}}, implemented as machine-based evaluation of match to cover minor spelling alterations or other edge cases \cite{roychowdhury2024evaluationragmetricsquestion}.
\end{itemize}

\subsection{Baselines: fully-cascaded and semi-cascaded RAG}
\label{subsec:baselines}
We implement two types of cascaded RAG baselines.
The \textit{fully-cascaded baseline} consists of a cascaded text retriever (an ASR module and a text retriever) and an LLM generator.
This framework transcribes the audio passages and treats spoken content retrieval as a text retrieval problem.
The \textit{semi-cascaded baseline} uses our speech retriever to retrieve the audio passages, then uses the transcripts of the audios and uses them to condition generation.
For our baselines, we use Qwen-7B-Chat as the text LLM generator.

To investigate the effect of transcription quality of the ASR step, we implement different versions of the baseline, each with
varying levels of average WER.
We use the transcriptions as ground truth text (where we assume that the WER is 0\%), while
the  case uses the same audio encoder used for the speech adapter.
To simulate a severe WER scenario, we use a small ASR module\footnote{\url{https://hf.co/jkang/espnet2\_librispeech\_100\_conformer\_word}} trained on 100hrs of Librispeech \cite{panayotov2015librispeech}.

%
Both baselines use the same frozen text retriever as the speech retriever and the text-only version of the SLM (Qwen-7B-Chat).


\section{Results}
\label{sec:results}


\subsection{Retrieval results}
\label{subsec:retrieval_results}

\begin{table}[!tbp]
\caption{Retrieval results. Passage WER reports the average WER of the text transcripts of the speech data.
The speech retriever operates directly on audio passages and achieves performance on par with retrieval on ground truth text (WER 0\%).
}
\centering
\resizebox{1.\linewidth}{!}{
\begin{tabular}{lcccc}
\toprule
                        & \textbf{Passage WER} & \textbf{Recall@5} & \textbf{Recall@10} & \textbf{Recall@100} \\
\midrule
\multicolumn{5}{c}{\textbf{SpokenSQuAD}}                                            \\ \midrule
GT Text Baseline  &       0\%       &   0.9707       &  0.9871         &   0.9985         \\ 
Low WER cascaded  &     $\sim$20\%         &  0.9525       &  0.9745        &   0.9974         \\
High WER cascaded &     $\sim$35\%         &  0.8768       &   0.9271      &    0.9926        \\
Speech Retriever  &      N/A        &  \textbf{0.9702}       &   \textbf{0.9869}     &   \textbf{ 0.9986}       \\ \midrule
\multicolumn{5}{c}{\textbf{VoxPopuli}}                                              \\ \midrule
GT Text Baseline      &      0\%        &    0.9942      &    0.9971      &   1.0         \\ 
Low WER cascaded        &     $\sim$17\%         &  0.9826       &   0.9855      &      0.9961      \\
High WER cascaded        &      $\sim$45\%        &   0.7106      & 0.7493       &   0.8858        \\
Speech Retriever  &       N/A      &   \textbf{0.9952}       &   \textbf{0.9981}   &   \textbf{0.9990}   \\ \bottomrule
\end{tabular}
}
\label{tab:retrieval_results}
\end{table}

We show our retrieval experiment results in \autoref{tab:retrieval_results}.
We observe that across both datasets, our proposed speech retriever outperforms the cascaded baselines.
This shows that our method offers a retrieval performance advantage even when the ASR step is done using a relatively high-performance ASR model. 
While the comparison between the  cascaded model and the ground truth text baseline indicates that the text retrieval model 
can be robust against low WER error rates,
the large performance drop in the High WER results shows that noisy transcriptions can lead to severe retrieval performance degradation.
Our speech retrieval removes the possibility of ASR error propagation at the retrieval step by operating directly on speech.

Moreover, our speech retriever achieves a retrieval performance that is on par with that of the ground truth transcript-based retrieval,
even obtaining a slightly higher performance in one metric (Recall@10, VoxPopuli set), showing that
our proposed speech retriever is a practical and powerful alternative to the cascading framework of spoken content retrieval.

\begin{figure}%
    \centering
    \subfloat[\centering Result on SpokenSQuAD]{{\includegraphics[width=0.45\textwidth,height=\textheight,keepaspectratio]{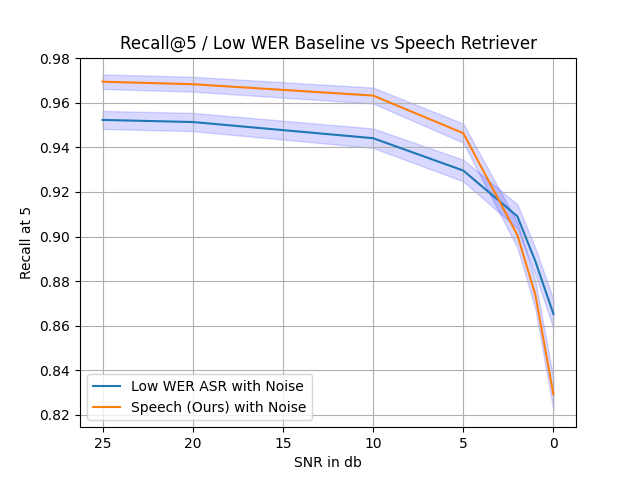} }}%
   \\
        \subfloat[\centering Result on VoxPopuli]{{\includegraphics[width=0.45\textwidth,height=\textheight,keepaspectratio]{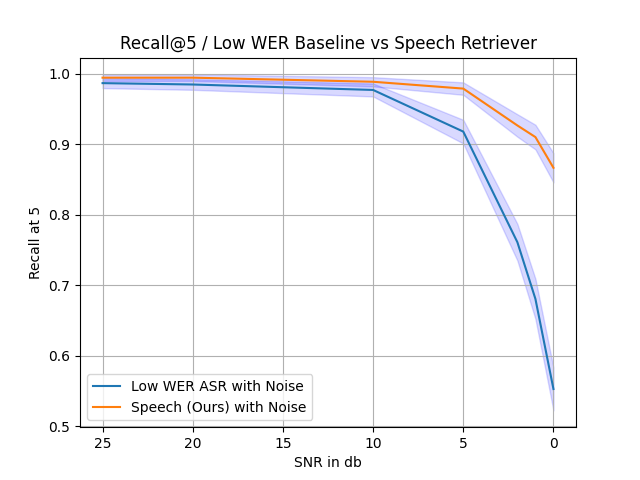} }}%
    \caption{
    Retrieval performance comparison with injected noise. We inject Gaussian noise to the speech signals and compare the text-based vs. our end-to-end retriever at different SNRs.
}%
    \label{fig:snr_plot}%
\end{figure}

\subsection{Retrieval performance under different noise levels}
\label{subsec:varied_noise_results}
To study the effect of noise on retrieval performance, we plot the Recall$@5$ of the text baselines and our speech retriever at
different noise levels, as shown in
\autoref{fig:snr_plot}. 
We add Gaussian noise at different signal-to-noise (SNR) levels to the audio.
We observe that across both datasets, our speech retriever is more robust to noise than the cascaded baseline that uses the same speech encoder as our end-to-end retriever, with the exception of a very high noise setting
for the SpokenSQuAD dataset (which is TTS-based).

\subsection{Generation results}
\label{subsec:generation_results}
Table~\ref{tab:generation_results} shows our comparison of SpeechRAG, fully-cascaded, and semi-cascaded RAG baselines.
Promisingly, we find that the full SpeechRAG framework outperforms the High-WER cascaded baselines on both datasets, highlighting the potential of SpeechRAG. 
For example, Table~\ref{fig:generation_example} shows how SpeechRAG avoids the corruption of context information by an inaccurate transcription of named entities.
However, we also find that it performs worse than the fully- and semi-cascaded baselines under settings, with a larger gap for the SpokenSQuAD dataset.
This is possibly due to the difference in the durations of the retrieved audios, with an average SpokenSQuAD audio lasting 4 times as long as the average VoxPopuli utterance.
This adversely affects performance since SLMs are typically not trained to handle multiple, long-context audios.

\begin{table}[!tbp]
\caption{Generation results. We use top-$5$ passages from the previous retrieval experiment as context.
SpeechRAG performs better than the baselines High WER cases, but not in  cases.
}
\centering
\resizebox{1.\linewidth}{!}{
\begin{tabular}{lccc}
\toprule
                              & \textbf{Passage WER}   & \textbf{Exact Match} & \textbf{LLM Correctness} \\ \midrule
\multicolumn{4}{c}{\textbf{SpokenSQuAD}}                                                         \\ \midrule
GT Text Baseline              &         0\%                   &    0.7514         &      0.8352           \\
\midrule
Fully-cascaded Low WER               &         $\sim$20\%                  &     0.5019      &     0.6987         \\
Fully-cascaded High WER              &          $\sim$35\%               &    0.2684         &       0.3701          \\
Semi-cascaded GT Text      &          0\%               &      \text{0.7364}       &       \text{0.8013}       \\
Semi-cascaded Low WER           &             $\sim$20\%                &     0.5057        &    0.7072           \\
Semi-cascaded High WER         &         $\sim$35\%                &   0.2787          &   0.3842           \\
\midrule
SpeechRAG    &       N/A               &   0.3522          &   0.4811           \\
\midrule
\multicolumn{4}{c}{\textbf{VoxPopuli}}                                                         \\ \midrule 
GT Text Baseline              &          0\%                  &      0.9080       &    0.9003             \\
\midrule
Fully-cascaded Low WER               &         $\sim$17\%                  &       0.7473    &       0.7561          \\
Fully-cascaded High WER              &             $\sim$45\%            &      0.4511      &   0.3950             \\
Semi-cascaded GT Text     &          0\%               &        \text{0.9158}     &     \text{0.9071}            \\
Semi-cascaded Low WER        &           $\sim$17\%               &         0.7301  &       0.7561        \\
Semi-cascaded High WER      &          $\sim$45\%               &       0.4327   &      0.3766          \\
\midrule
SpeechRAG       &       N/A               &      0.8045       &     0.7173         \\
\bottomrule
\end{tabular}
}
\label{tab:generation_results}
\end{table}

\begin{figure}[!t]
\begin{tcolorbox}[
    before upper={\setlength{\parskip}{0.5em}},
    width=\columnwidth
]
    \textbf{Passage:} \textit{...whilst under Soviet rule, Armenian classical music composer \textbf{Aram Khatchaturian} became internationally well known for his music, for various ballets and the Sabre Dance...}
    
    \textbf{Text query:} \textit{Who composed the Sabre Dance?}
    
    \textbf{Ground truth answer:} \textit{\textbf{Aram Khatchaturian}}
    
    \textbf{ASR transcript (top-1 retrieved):} \textit{… whilst under soviet rule armenian classical music composer \textbf{\color{red}aram cocheterien} became internationally well known for his music for various ballets and the sabre dance …}
    
    \color{red}{\textbf{Fully-cascaded generation:} \textit{\textbf{Aram Cocheterien} composed the Sabre Dance.}}
    
    \color{ForestGreen}{\textbf{SpeechRAG generation:}~\textit{The Sabre Dance was composed by \textbf{Aram Khachaturian}.}}
    \end{tcolorbox}
    \caption{SpokenSQuAD generations of the fully-cascaded model vs our ASR-less SpeechRAG framework. The named entity transcription error in the ASR step propagates to the generation step, while the SLM of SpeechRAG correctly generates the named entity.}
    \label{fig:generation_example}
\end{figure}

\section{Conclusion}
\label{sec:conclusion}
In this work, we propose a first fully speech-based solution to question answering over speech.
To achieve this, we implement a speech retriever consisting of a speech adapter and a frozen LLM-based text retriever and 
show that by indexing and retrieving speech directly, our framework outperforms 
cascaded retrieval in noisy ASR scenarios, and matches ground truth text retrieval.
In the generation step, our framework bypasses ASR by using an SLM conditioned on retrieved audio.
While the SpeechRAG generation outperforms cascaded baselines in high WER scenarios, we identify the potential for improvement in the performance gap between our framework and the low WER baseline generations.


\bibliographystyle{IEEEbib}
\bibliography{refs}
\label{sec:refs}

\end{document}